\documentclass[twocolumn,amsmath,amssymb,showpacs,preprintnumbers]{revtex4}
\usepackage{dcolumn}
\usepackage{bm,amsmath,graphicx}

\begin{document}

\newcommand{\het}{$^{3}$He{ }}
\newcommand{\ayo}{A$_{y0}$}
\newcommand{\aoy}{A$_{0y}$}
\newcommand{\ayy}{A$_{yy}$}
\newcommand{\axx}{A$_{xx}$}

\title{Spin-Correlation Coefficients and Phase-Shift Analysis for p+\het Elastic Scattering}
\author{T. V. Daniels\footnote{Electronic address: tvdaniels@physics.umass.edu; present address: Department of Physics, University of Massachusetts, Amherst, MA 01003}, C. W. Arnold, J. M. Cesaratto, T. B. Clegg, A. H. Couture, H. J. Karwowski, and T. Katabuchi\footnote{Present address: Research Laboratory for Nuclear Reactors,
Tokyo Institute of Technology,
Ookayama, Meguro-ku, Tokyo 152-8550,
JAPAN} }
\affiliation{The  University of North Carolina at Chapel Hill, Chapel 
Hill, North Carolina 27599-3255, USA\\
Triangle Universities Nuclear Laboratory, Durham, North Carolina 
27708-0308, USA}
\date{\today}
\begin{abstract}
Angular distributions for the target spin-dependent observables \aoy, \axx, and \ayy{} have been measured using polarized proton beams at several energies between 2 and 6  MeV and a spin-exchange optical pumping polarized \het target.  These measurements have been included in a global phase-shift analysis following that of George and Knutson, who reported two best-fit phase-shift solutions to the previous global p+\het elastic scattering database below 12 MeV.  These new measurements, along with measurements of cross-section and beam-analyzing power made over a similar energy range by Fisher \textit{et al.}, allowed a single, unique solution to be obtained.  The new measurements and phase-shifts are compared with theoretical calculations using realistic nucleon-nucleon potential models.

\end{abstract}
\pacs{21.45.-v, 21.30.-x, 24.70.+s, 25.40.Cm}
\maketitle

\section{Introduction}

Beginning from the picture that atomic nuclei are composed of interacting nucleons, \textit{ab initio} calculations of light nuclear systems are based on realistic nucleon-nucleon potential models, which have been adjusted to reproduce two-nucleon (NN) scattering and bound-state data accurately \cite{Car98}.  This effort has included calculations of low-energy scattering observables for the three-nucleon (3N) and four-nucleon (4N) systems.  The latter calculations are especially significant because the 4N system is the lightest to exhibit thresholds and resonances \cite{Til92}, so that its correct description is an important milestone for this approach. 

The comparison of \textit{ab initio} calculations with nucleon-deuteron scattering measurements reveals general agreement for the cross-section and tensor analyzing powers, but significant underprediction of the beam and target analyzing powers \cite{Glo96}.  A similar ``A$_y$ Puzzle'' has been reported for p+\het elastic scattering \cite{Fis06}.  All NN models and theoretical methods yield this disagreement, which is not resolved by including the 3N force necessary to reproduce the 3N and 4N binding energies \cite{Kie10}.

The study of this discrepancy may benefit from the more thorough comparison between theory and experiment made possible by a set of experimental phase-shifts and mixing parameters.  For example, A$_y$ is known to be particularly sensitive to the splitting between triplet P-wave phase-shifts \cite{Fis06}.  A wealth of experimental data exists for p+\het elastic scattering below 12 MeV proton energy, with the most recent phase-shift analysis by George and Knutson \cite{Geo03} performed on a database of over 1000 data points.  That analysis, however, was unable to constrain a unique set of parameters, and instead obtained two solutions which fit the data equally well.  The difference between the two solutions was largest for spin-correlation coefficients below 4 MeV, where no such data exist.  With the aim of resolving the phase-shift ambiguity, we have used a polarized \het target \cite{Kat05} to measure angular distributions of the spin-correlation coefficients \axx{} and \ayy{} at proton energies between 2 and 6 MeV and included those new data, along with those of Fisher \textit{et al.} \cite{Fis06}, in a new global phase-shift analysis following that of George and Knutson.

\section{Experiment}

\subsection{Polarized Target}
Polarized and unpolarized beams from the Triangle Universities Nuclear Laboratory (TUNL) tandem accelerator were directed by an analyzing magnet to a 62 cm diameter scattering chamber.  The polarized \het target was specifically designed for low-energy charged-particle scattering experiments and has been described previously in detail \cite{Kat05}.  In contrast to previous targets used for the same purpose \cite{Roh68}-\cite{All93a} which operated by metastability-exchange optical pumping (MEOP), the present target used spin-exchange optical pumping (SEOP).  This method has previously been used in polarized \het targets for electron \cite{Ant96} and gamma-ray \cite{Kra07} scattering experiments.  

The primary advantage of SEOP over MEOP is greater target thickness, since it polarizes \het at a pressure of about 8 bar instead of several mbar.  The need to minimize energy loss for the incident and scattered particles for the low-energy application, however, required the use of thin windows to contain the gas.  Since such windows cannot withstand the full 8 bar \het pressure, we optically pumped $^3$He in a system separate from the scattering target cell.  The target cell was then batch-filled with polarized gas to a pressure of approximately 1 bar. 

The target cell was a 5.1 cm Pyrex sphere with openings along the equator for the incident and scattered particles.  These were covered with 7.6 $\mu$m Kapton foil affixed with Torr Seal epoxy \cite{Var}.  The cell was housed in a compact sine-theta coil to provide a uniform 0.7 mT magnetic holding field.  An NMR coil pressed against the rear of the cell was used to measure the target polarization, as discussed below. 

The \het polarization was produced by SEOP using Rb as the intermediate alkali metal.  A 60W fiber-coupled diode laser system tuned to the 795nm Rb D1 absorption line provided, with appropriate optics, the circularly polarized light for the optical pumping.  Two modifications to our original polarizer \cite{Kat05} were attempted.  In agreement with results reported by others \cite{Bab03}, the use of ``mixed metal'' optical pumping cells containing both Rb and K was found to decrease the ``spin-up'' time \cite{Cou06}.  In this work, the typical time required to reach saturation polarization was about 12 h with mixed-metal cells, compared to about 24 h using only Rb.  

The other modification was the use of frequency-narrowed laser light for optical pumping.  Following the work of \cite{Cha03}, an external Littrow cavity was constructed and used to reduce the output width of 50 W Quintessence \cite{Qui} diode bar array from 2 nm to 0.3 nm \cite{Arn06a}.  The narrowed output power was about 30 W.  Unfortunately, the $^3$He polarization produced with this laser system was not consistently higher than that produced with the 60-80 W broadband system, perhaps because more light was absorbed from the latter.  The majority of the present data was therefore taken with the broadband laser.

\subsection{Scattering Measurements}
\begin{figure*}
\begin{center}
\includegraphics[width=6in]{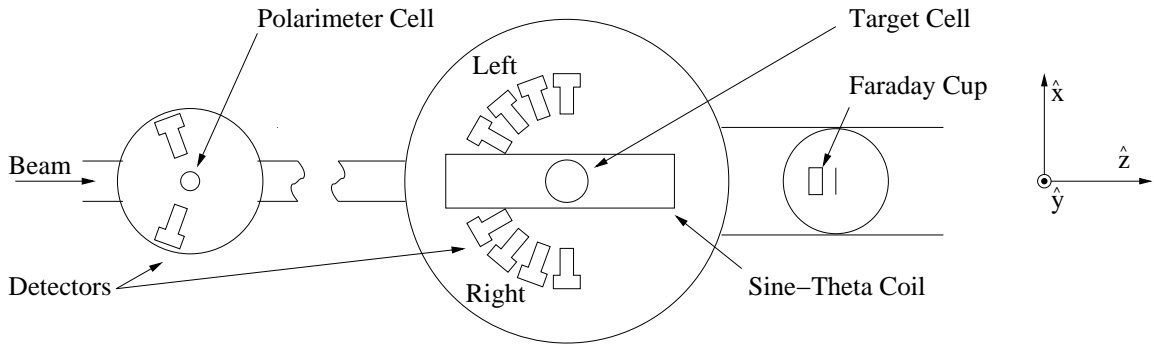}
\caption{Diagram of the experimental arrangement.  The polarimeter chamber at left was used with the detectors either horizontal, as shown, or vertical.  The polarimeter cell was removed from the beam during data taking to allow beam to reach the target chamber.  The coordinate system used to define the scattering observables is also shown. }\label{chamber}
\end{center}
\end{figure*}

The experimental arrangement is illustrated in Fig. \ref{chamber}.  Beam current on the target cell was limited to 50 nA to minimize damage to the Kapton foils.  Failure of the epoxy, especially that sealing the beam-exit foil, caused cells to leak after a few days.  Each such failure required a cell change and recalibration of the NMR signal (see below).

Measurements were made at five proton energies below 6 MeV which overlap the energies of both Fisher \textit{et al.} \cite{Fis06} and Alley and Knutson \cite{All93a}.  Analyzing magnet settings which determined the beam energy were selected according to the calibration of Ref. \cite{Fis03}.  Beam energies were adjusted to offset energy loss in the foils and gas, as modeled with TRIM \cite{Sri08}.  Bombarding energies for data taken with different thickness entrance foils were slightly different, and the error-weighted average value was adopted.  The uncertainty in TRIM stopping powers for materials used was estimated by comparison with experimental stopping powers \cite{Sri08}, and ranged from 3-10\%. An uncertainty of 10\% was assigned to cases where no data were present in the relevant energy range.

The beam and target polarizations were reversed frequently during data-taking.  The beam polarization was reversed at either 1 or 10 Hz in the sequence ``udduduud'', where ``u'' means ``spin-up" and ``d'' means ``spin-down''.  The target polarization was reversed less frequently, since a few seconds were required to reverse the target's magnetic field.  Polarized target data were collected for intervals of 2.5 m in each spin orientation, with NMR measurements of the magnitude of the polarization made immediately before and after the orientation was reversed.  The target polarization decayed with a 2-3 h time constant, so this process was stopped when the gas was judged to be too depolarized, generally after about 1 h.  At that time the gas was exhausted from the target cell, which was then flushed with research-grade N$_2$ and refilled with a new batch of polarized gas.  The recovered depolarized $^3$He gas was circulated through a LN$_2$-cooled trap to remove impurities before repolarization.

Scattered particles emerging from the target were detected by four pairs of Si detectors which could be rotated to the desired angle.  Available angles were restricted by the windows in the sine-theta coil's mu-metal shield to 20$^\circ$ increments between 30 and 150$^\circ$.  The shield could be moved axially so that ``intermediate'' angles offset by 10$^\circ$ were also available.  The detectors were each placed in an Al holder behind two brass collimators spaced 5.08 cm apart in an Al ``snout'', which restricted the range of scattering angles visible to the detector to 1.5$^\circ$.  Detectors were as close as possible to the target without the 30$^\circ$ snout's touching the sine-theta coil, so that the distance from the center of the target cell to the front collimators was about 10.2 cm.  Beam current on target was measured by a electrostatically suppressed Faraday cup located about 0.5 m behind the target cell.  Charge went to ground through a current integrator to measure the relative number of beam particles in each spin state.

\begin{figure}
\begin{center}
\includegraphics[width= 2.8in]{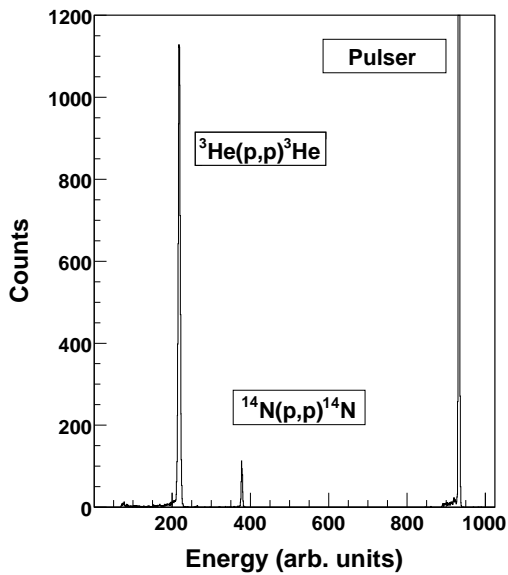}
\caption{ $^3$He(p,p)$^3$He spectrum taken at 5.54 MeV and 90$^\circ$.  } \label{spectrum}
\end{center}
\end{figure}

An example detector spectrum is shown in Fig. \ref{spectrum}.  The peaks corresponding to protons elastically scattered from \het and small amounts of N$_2$ were well separated.  The latter was required by the optical pumping process \cite{Kat05}. Occasionally $^4$He was added to the cell to measure the beam polarization, and in some cases that peak overlapped the \het peak.  In such cases the peaks were fit with skewed Gaussians.  A pulser was added to each spectrum to measure electronic dead time, which was typically less than 1\%. 

The observables were extracted from peak yields in left (L) and right (R) detectors using an extension of the geometrical mean method \cite{Ohl73} for analyzing powers to include polarized beam and target.   With the spins aligned vertically along the $\pm$y-axis, the following cross-ratios were formed:
\begin{align*}
X_1 &= \sqrt{\left(\frac{L\uparrow\uparrow + L\uparrow\downarrow}{L\downarrow\uparrow + L\downarrow\downarrow}\right)\left(\frac{R\downarrow\uparrow + R\downarrow\downarrow}{R\uparrow\uparrow + R\uparrow\downarrow }\right)} 
= \frac{1+p_yA_{y0}}{1-p_yA_{y0}} \nonumber\\
\nonumber\\ 
X_2 &= \sqrt{\left(\frac{L\uparrow\uparrow + L\downarrow\uparrow}{L\uparrow\downarrow + L\downarrow\downarrow}\right)\left(\frac{R\uparrow\downarrow + R\downarrow\downarrow}{R\uparrow\uparrow + R\downarrow\uparrow }\right)} 
= \frac{1+p^T_yA_{0y}}{1-p^T_yA_{0y}}\nonumber\\
\nonumber\\ 
X_3 &= \sqrt{\left(\frac{L\uparrow\uparrow + L\downarrow\downarrow}{L\uparrow\downarrow + L\downarrow\uparrow}\right)\left(\frac{R\uparrow\uparrow + R\downarrow\downarrow}{R\uparrow\downarrow + R\downarrow\uparrow }\right)} 
= \frac{1+p_yp^T_yA_{yy}}{1-p_yp^T_yA_{yy}},\nonumber
\end{align*}
where the arrows indicate the beam and target spin state, for example $L\uparrow\downarrow$ refers to the number of particles scattered into the left detector while the beam was spin ``up'' and the target was spin ``down''.  The detector yields were normalized to the current integration and target pressure, which was known to about 2\%, for each spin state.  The polarization of the beam and target are given by $p$ and $p^T$, respectively.  The observables are therefore
\begin{align} 
A_{y0} &= \frac{1}{p_y}\left(\frac{X_1 - 1}{X_1 + 1}\right)  \\ \label{yobs} 
A_{0y} &= \frac{1}{p^T_y}\left(\frac{X_2 - 1}{X_2 + 1}\right)\\  
A_{yy} &= \frac{1}{p_yp^T_y}\left(\frac{X_3 - 1}{X_3 + 1}\right).
\end{align}
Similarly, when the beam and target spins are aligned horizontally along the x-axis,
\begin{align} \label{xobs}
A_{xx} &= \frac{1}{p_xp^T_x}\frac{X_3 - 1}{X_3 + 1}.
\end{align}

If either the beam or target is unpolarized, only one analyzing power will be non-zero, and its expression reduces to the usual cross-ratio for analyzing powers,                                                                                                         \begin{align}\label{unpol}
A_{y} = \frac{\sqrt{\frac{L\uparrow R\downarrow}{L\downarrow R\uparrow}} - 1}{\sqrt{\frac{L\uparrow R\downarrow}{L\downarrow R\uparrow}} + 1}.
\end{align}
The same is true for the scattering of spin-1/2 protons from spin-0 $\alpha$-particles used for beam polarimetry.

\subsection{Beam Polarimetry}
Proton beam leaving an atomic beam polarized ion source \cite{Cle95} passed through a calibrated Wien filter at the ion source to orient the spin quantization axis of the beam in the desired direction at the scattering chambers. The magnitude of the beam polarization was measured periodically using p+$^4$He elastic scattering in either the target cell or in a separate cell in a polarimeter chamber installed upstream of the target chamber.  Detectors in the latter could be mounted at 110$^\circ$ in the horizontal or vertical scattering planes, so that either polarization component could be measured. The cell in the polarimeter chamber could be moved to insert it periodically into the beam for polarization measurements.  
 
The p+$^4$He asymmetries were divided by the analyzing power A$_y$ to obtain the beam polarization.  Published phase shifts \cite{Sch71} were used in a
spin$\frac{1}{2}$-on-spin-0 phase-shift code to calculate the analyzing powers for the energies at the center of the cell as determined from TRIM.  The uncertainty in the resulting analyzing power was typically 2\%.

For more than half of the spin-correlation data, however, the beam polarization was unstable, so that periodic monitoring did not necessarily determine the average polarization.  Therefore, the beam polarization for all A$_{yy}$ measurements was determined by normalizing our relative A$_{y0}$ measurements to published values \cite{Fis03, All93a}.  Each point in a relative A$_{y0}$ angle set was divided by a value linearly interpolated from those previous measurements at the same energy, and the polarization was taken to be the average of these ratios.  An uncertainty of 0.02 was assigned to the polarization and added in quadrature with statistical uncertainties.  No published data were available at 2.7 MeV, so smooth curves were fit through existing distributions at each angle vs. energy and evaluated at 2.7 MeV.  The normalization then proceeded as above.

The procedure was extended to about one-third of the A$_{xx}$ measurements by ``tipping'' the spin 20$^\circ$  out of the plane with the Wien filter and applying the above analysis to the y-component.  The x-component of the beam polarization was obtained by multiplying the y-component by the ratio of the two components.  An 8\% normalization uncertainty was applied to those angular distributions to account for the estimated 1.5$^\circ$ uncertainty in the relative azimuthal orientation of the scattering plane and the polarized beamÕs quantization axis.  The remaining A$_{xx}$ measurements with stable beam polarization relied on polarimeter measurements as described above.

\subsection{Target Polarimetry}
As discussed in detail by Katabuchi \textit{et al.} \cite{Kat05}, the target polarization was monitored using pulsed NMR.  Briefly, an RF pulse at the Larmour frequency set by the sine-theta coil magnetic field was sent through a small coil pressed against the rear of the target cell.  The resulting collective precession of $^3$He spins about that magnetic field induced a signal proportional both to the $^3$He polarization and pressure. This voltage was then divided by the cell pressure to give a relative measure of polarization.  

These relative NMR data for each target cell were calibrated against separate $^4$He + $^3$He A$_{y}$ measurements at an energy and angle where A$_{y}$= -1.  The resulting scattering asymmetries for an incident $^4$He beam, given by Eq. \ref{unpol}, were therefore direct measurements of the $^3$He target polarization.  This calibration method was motivated by the prediction of Plattner and Bacher \cite{Pla71} of an A$_y$ =  -1 extremum near 15.33 MeV $^3$He lab energy and 47$^\circ$ $^3$He lab scattering angle.  Their prediction for its location was only approximate.  Thus, \textit{relative} measurements of A$_y$ in $^4$He+$^3$He elastic scattering as a function of angle and energy near the predicted extremum were made to define the local minimum.  We determined $\theta$$_{min}$ to be 46.64 $\pm$ 0.22$^\circ$, and the measurements of A$_y$ vs energy agree with the prediction of Ref. \cite{Har70} of a very broad minimum.

The simultaneous NMR-scattering calibrations for various target cells were made at 45$^\circ$, and E$_\alpha$ between 15.44 and 15.82 MeV.  Since this was not exactly the minimum point for A$_y$, a target polarization value was assigned from the relative measurements with an uncertainty of 3\%, normalized so that the minimum A$_y$ was equal to -1.  

\subsection{Steering Effect}

\begin{figure}
\begin{center}
\includegraphics[width = 3in]{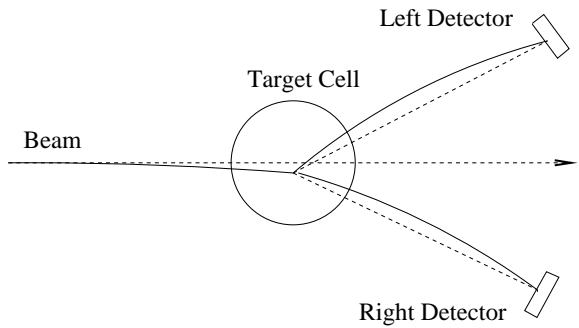}
\caption[Steering of the incoming proton beam and scattered particles by the sine-theta coil magnetic field]{Steering of the incoming proton beam and scattered particles by the sine-theta coil magnetic field.  The target cell and detectors are shown from above.  The beam is incident from the left on the gas cell along the dotted line, but deflected as shown by the magnetic field, which is oriented out of the page.   Similarly, the scattered particles travel along curved paths to the detectors.  The figure is not to scale, and the size of the effect is exaggerated for clarity.}\label{steer}
\end{center}
\end{figure}

The target magnetic field, though small, steered the incoming and scattered protons slightly through the Lorentz force, as sketched in Fig. \ref{steer}.  When the B-field was reversed to reverse the target spin, the particles were steered oppositely, and the relationship between detector yields reversed.  

First, the incoming beam deflection moved the scattering center closer to one detector and further from the other, thus changing their relative solid angles.  Second, the scattered protons were deflected to emerge from the target cell at slightly different angles than those at which they were actually scattered.  Thus, for a given orientation of the magnetic field, the actual scattering angle of particles reaching one detector of a left/right pair was more forward than the detector angle, while that of those reaching the other detector was more backward. This difference in scattering angle produced an instrumental asymmetry through the angular dependence of the differential cross section. These two steering effects, unlike other systematic effects \cite{Ohl72}, produced systematic instrumental asymmetries which were \textit{not} cancelled by reversing the target spin. 

\begin{figure}
\begin{center}
\includegraphics[width = 3in]{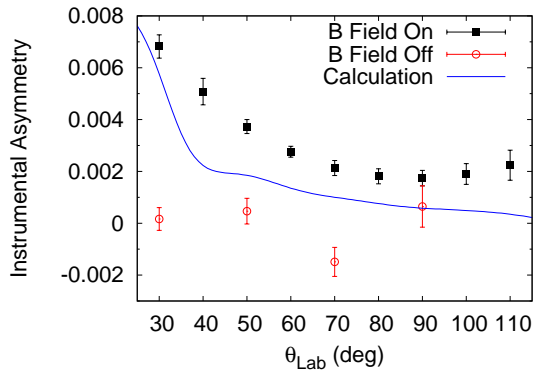}
\caption{(Color online) Instrumental target asymmetries measured at 2.25 MeV with the sine-theta coil magnetic field both on and off.  A simple calculation of the effect of magnetic steering is also shown.}\label{effect}
\end{center}
\end{figure}

These effects were largest at our lowest bombarding energies. Fig. \ref{effect} shows observables extracted from data taken while the target was unpolarized, when the asymmetries corresponding to the target analyzing power and spin-correlation coefficients should be zero.  While this was true for A$_{0y}$ measured with the sine-theta coil's magnetic field turned off, non-zero asymmetries were obtained with the field on.  The asymmetries for A$_{yy}$ and A$_{xx}$, on the other hand, were consistent with zero even with the field on. 

The result of a simple calculation of these two effects at 2.25 MeV is also shown in Fig. \ref{effect}.  Though the largest deflection angle calculated was 0.1$^\circ$, the resulting asymmetries are large enough to interfere with polarized target measurements.  The calculation neglects the finite size of the beam and target, and simply determines the energy loss, modeled in TRIM, and magnetic steering of incoming and scattered protons in small steps as they proceed through the magnetic field.  The calculation reproduces the general size and forward-angle trend of the effect, but with insufficient detail to be used to correct the data.  

Instead, actual measurements of these instrumental asymmetries, obtained both from direct measurements with unpolarized $^3$He and by extrapolating polarized target asymmetries to zero polarization, were subtracted from polarized target A$_{0y}$ data.  The corrections obtained in this way were often large, being several times the size of the observable at forward angles at the lowest proton energies.

\subsection{Results for Observables}

\begin{figure}
\begin{center}
\includegraphics[width= 3.5in]{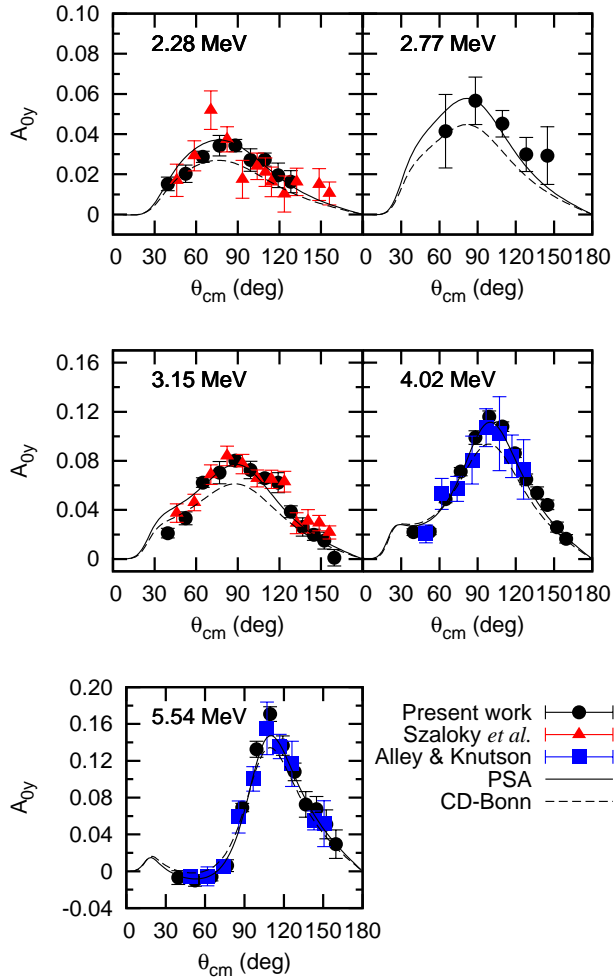}
\caption{(Color online) Present measurements of A$_{0y}$ at five proton beam energies, together with the global PSA fit.  Measurements of Szaloky \textit {et al.} \cite{Sza78a} and Alley and Knutson \cite{All93a} as well as theoretical calculations of Deltuva \cite{Del07b} using the CD-Bonn potential \cite{Mac01} are also shown. }
\label{a0y}
\end{center}
\end{figure}

\begin{figure}
\begin{center}
\includegraphics[width= 3.5in]{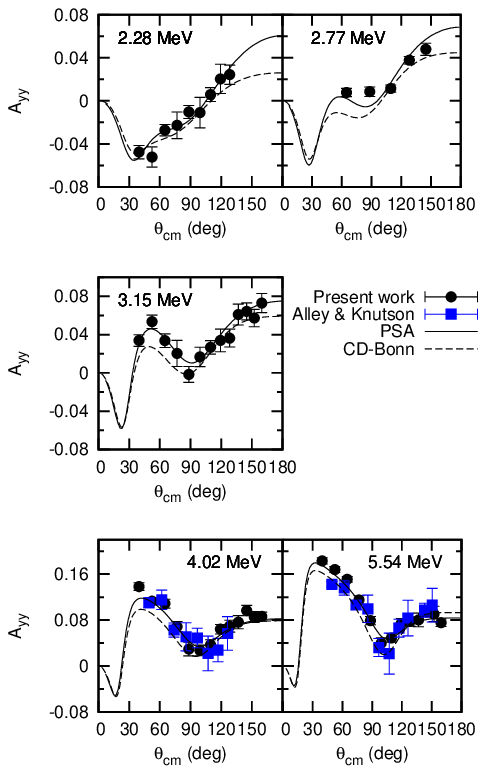}
\caption{(Color online) Present measurements of A$_{yy}$ at five proton beam energies, together with other results as mentioned for Fig. 8.}
\label{ayy}
\end{center}
\end{figure}

\begin{figure}
\begin{center}
\includegraphics[width= 3.5in]{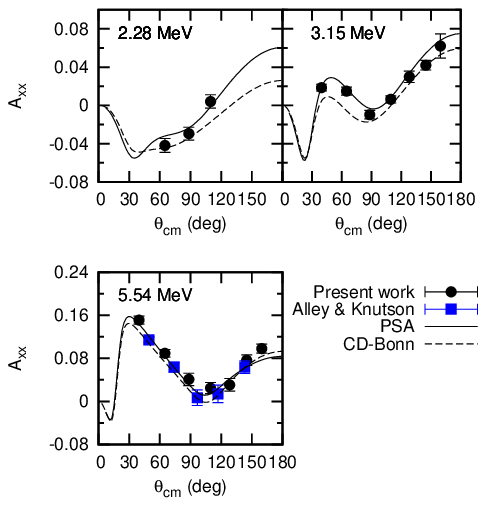}
\caption{(Color online) Present measurements of A$_{xx}$ at three proton beam energies, together with other results as mentioned for Fig. 8. }
\label{axx}
\end{center}
\end{figure}

The present measurements of A$_{0y}$, A$_{yy}$, and A$_{xx}$ are shown in Figs. \ref{a0y}-\ref{axx}.  Each plot also includes a curve calculated from the new best-fit effective range parameters of the global phase-shift analysis (PSA) discussed below, as well as previous measurements where available.  The overall agreement with the previous measurements is good, with the most forward-angle A$_{yy}$ points of Alley and Knutson \cite{All93a} at 5.54 MeV being the only exception, and the present results have smaller error-bars.  The new measurements are well-fit by the phase-shift analysis, except for the two most forward-angle A$_{0y}$ points at 3.15 and 4.02 MeV.  The forward-angle points required the largest correction for magnetic steering, so the disagreement with the PSA may indicate that the correction applied to those points was not sufficiently accurate.  

\section{Phase-Shift Analysis}
A phase-shift analysis of the global p+$^3$He elastic scattering database below E$_p$= 12 MeV was performed following the earlier work of George and Knutson \cite{Geo03}, with the addition of about 300 new data points, including the $\frac{d\sigma}{d\Omega}$ and A$_{y0}$ measurements of Fisher \textit{et al.} \cite{Fis06} and the present A$_{0y}$, A$_{yy}$, and A$_{xx}$ measurements.  These additional data all fell between 1.0 and 5.54 MeV.   The search routine was the same as that used in the previous analysis and was provided by George \cite{Geo06}.  

The program calculated scattering observables as functions of scattering matrix elements, which were in turn parameterized using phase-shifts and mixing parameters according to the Blatt-Biedenharn convention \cite{Bla52}.  The phase-shifts and mixing parameters used were 
$^1$S$_0$, $^3$S$_1$, $^1$P$_0$, $^3$P$_2$, $^3$P$_1$, $^3$P$_0$, $^1$D$_2$, $\epsilon\left(1^-\right)$, $\epsilon\left(1^+\right)$,  and $\epsilon\left(2^-\right)$, as well as consolidated $^3$D$_j$ and $^3$F$_j$ triplet phase-shifts.  The energy dependence of the phase-shifts and mixing parameters was described by the first three terms in a modified effective-range expansion.  These 36 effective range parameters were adjusted to minimize $\chi^2$ with respect to the experimental database using the MINUIT package \cite{Jam94}.  As described in Ref. \cite{Geo03}, the database was broken into groups of measurements thought to have common normalizations.  These 21 normalization factors were analytically adjusted at each step of the parameter search to further minimize $\chi^2$. 

Initial parameter searches resulted in multiple solutions which were discontinuous in one or more of the $^1$D$_2$, $^3$D$_j$, and $^3$F$_j$ phase-shifts.  The discontinuity, discussed also by Alley \cite{All92}, occurred when the phase-shift crossed zero and was deemed unphysical.  The number of such solutions was reduced by fixing the small $^3$F$_j$ phase-shift at the values obtained using the database of Alley and Knutson \cite{All93a}.  The best-fit $\chi^2$ when this parameter was not searched increased by only 0.2\%.  All but one of the remaining solutions, which spanned a range of about 10\% in $\chi^2$, were rejected by demanding that the phase-shifts be continuous in the energy range covered by the database.  The remaining solution, which has the lowest $\chi^2$ and is adopted as the present global result, yields a small positive scattering length for $^3$D$_j$, indicating that a discontinuity must occur in that phase at an energy below the lower end of the database (100 keV).  The best-fit effective range parameters and associated statistical uncertainties are given in Table \ref{effective_range}.

\begin{table} 
\begin{center}
\caption{Best-fit effective-range parameters with statistical uncertainties.} 
\begin{tabular}{ |c|r@{$\pm$}l | r@{$\pm$}l | r@{$\pm$}l |}
\hline
Phase & \multicolumn{2}{c|}{a0$\times10^{-2}$ } & \multicolumn{2}{c|}{a1$\times10^{-1}$} & \multicolumn{2}{c|}{a2} \\ \hline
$^1$S$_0$ &  -9.0 & 0.4 & 7.9 & 0.6 & 1.0 & 0.2 \\ \hline 
$^3$S$_1$ &  -11.06 & 0.17 & 7.5 & 0.3 & -0.09 &0.08 \\ \hline 
$^1$P$_1$ &  5.44 & 0.15 & -1.7 & 0.3 & 4.05 & 0.13 \\ \hline 
$^3$P$_2$ &  2.128 & 0.012 & 1.591 & 0.018 & 0.356 & 0.009 \\ \hline 
$^3$P$_1$ &  1.63 & 0.02 & 2.06 & 0.04 & 0.592 & 0.018 \\ \hline 
$^3$P$_0$ &  8.8 & 0.3 & -1.5 & 0.3 & 2.40 & 0.09 \\ \hline 
$^1$D$_2$ &  -14 & 4 & 15 & 4 & -8.3 &1.3 \\ \hline 
$^3$D$_j$ &  -0.06 & 0.19 & -3.4 & 0.9 & 7.2 & 0.4 \\ \hline 
$^3$F$_j$ &  \multicolumn{2}{c|}{4.19} & \multicolumn{2}{c|}{38.3} & \multicolumn{2}{c|}{-6.29} \\ \hline 
$\epsilon\left(1^+\right)$ &  -5 & 5 & 34 & 6 & -10.7 &1.7 \\ \hline 
$\epsilon\left(1^-\right)$ &  -420 & 11 & 230 & 10 & -41 & 2 \\ \hline 
$\epsilon\left(2^-\right)$ &  17 & 7 & 8 & 6 & -5.9 & 1.4 \\ \hline 
\end{tabular}
\label{effective_range}
\end{center}
\end{table}

The global solution had a $\chi^2$-per-datum of about 2 for the data added in this analysis.  This could be improved to between 1.3 and 1.5 if points whose individual $\chi^2$ contributions exceeded 10 were rejected.  About half of these 13 out of about 300 new points seemed simply to be random outliers, while the others seemed to be associated with systematic problems.  These included some forward angle A$_{0y}$ points which had been corrected for magnetic steering.  Another apparent systematic problem was found for A$_{y0}$ data at 1.60 MeV for which the four-most-forward angle points disagreed with the phase-shift analysis.  The effect of their removal on the phase-shifts was generally negligible, and in all cases within the range of systematic error indicated by the single-energy analyses described below.

In order to gauge the effects of systematic errors, single-energy analyses were performed at energies where new spin-correlation and new or existing cross-section measurements were available, \textit{i.e.} at nominal proton energies of 2.25, 3.13, 4.00, and 5.54 MeV.  All measurements within 100 keV of the nominal energies were included.  The same method was used for these single-energy fits as for the energy-dependent work, except that the phase-shifts were searched directly, instead of through the effective range parameters.  

\begin{figure*}
\begin{center}
\includegraphics[width=5.4in]{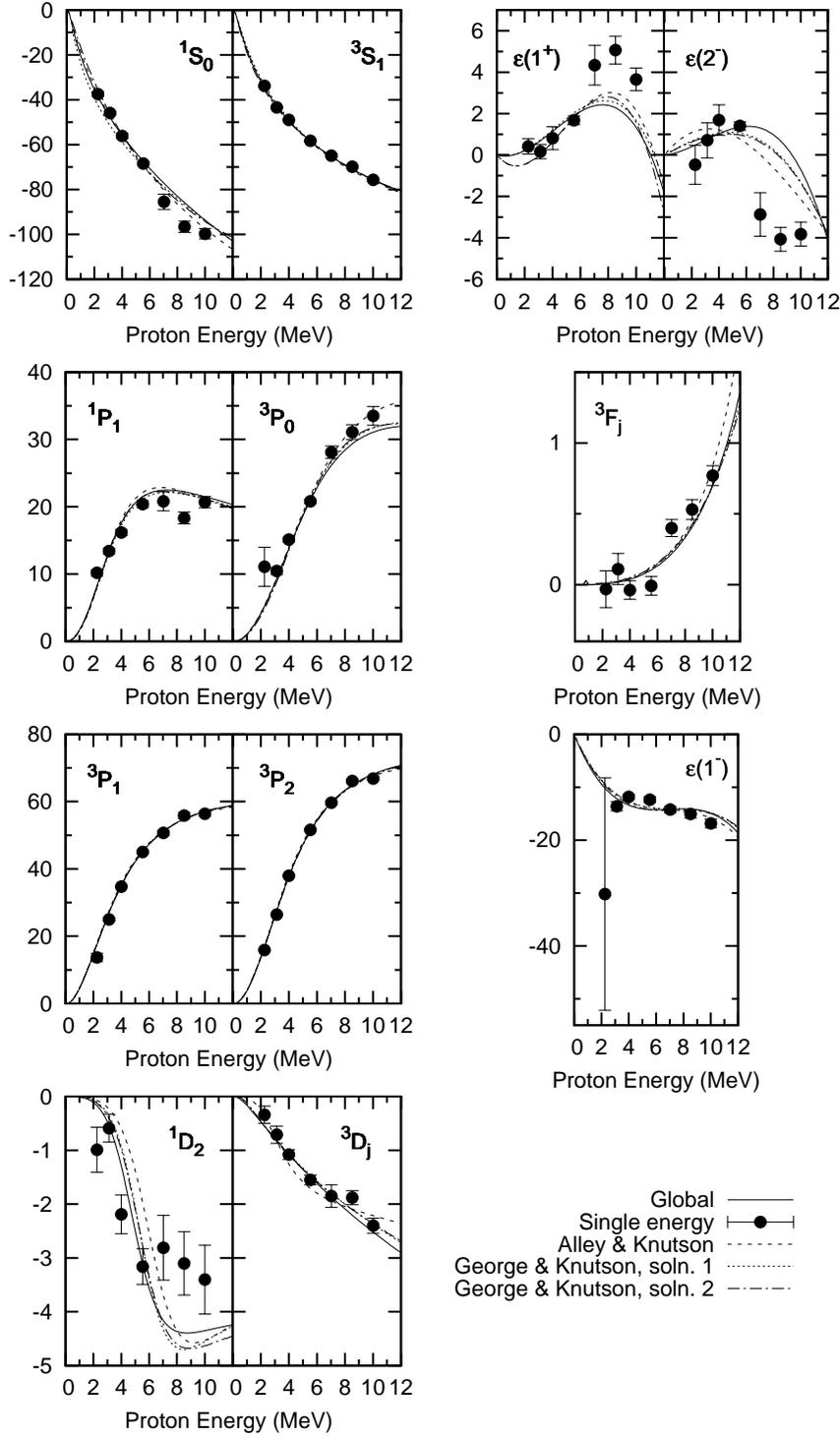}
\caption{Phase-shift results, in degrees, as functions of proton lab energy in MeV.  Both the global and single-energy results of this analysis are shown, as well as the previous results of Alley and Knutson \cite{All93b} and George and Knutson \cite{Geo03}. }
\label{phases}
\end{center}
\end{figure*}

The present phase-shift results, both single-energy and global, are shown in Fig. \ref{phases}, along with those of Refs. \cite{All93b} and \cite{Geo03}.  The addition of the new data removes the S-wave ambiguity in the latter results without qualitatively modifying the behavior of the other parameters, such as the resonant P-wave behavior associated with excited states of $^4$Li.   The new, low energy data also seem to introduce some tension with previous, higher energy data, as indicated by differences between the present global results and those of Ref. \cite{All93b} for $^1$S$_0$ and $^3$P$_0$.  

Results for selected phase-shifts, in degrees, are tabulated in Tables \ref{globalphasetable}-\ref{globalphasetable2} at the nominal energies of the single-energy analyses.  The quoted uncertainties include statistical uncertainties and systematic sources including differences between the single-energy and global analyses and variation of the parameters when outliers are excluded.  Theoretical calculations, described below, using the CD-Bonn realistic NN potential are also shown for comparison.

\begin{table} 
\begin{center}
\caption{Global phase-shift analysis results at 2.25 and 3.15 MeV} 
\begin{tabular}{ | c | r@{$\pm$}l | c | r@{$\pm$}l | c |}
\hline
 & \multicolumn{3}{c |}{2.25 MeV}& \multicolumn{3}{c |}{3.15 MeV} \\ \hline
{Phase} & \multicolumn{2}{c|}{Present} & {CD-Bonn}  & \multicolumn{2}{c|}{Present} & {CD-Bonn}\\ \hline
$^1$S$_0$ &  -39.1 &1.7 & -39.6 & -48.7 & 0.9 & -49.3\\ \hline 
$^3$S$_1$ &  -34.5 & 0.7 & -34.8 & -42.90 & 0.09 & -42.9\\ \hline 
$^1$P$_1$ &  8 & 2 & 10.6 & 13.4 & 0.4 & 14.9\\ \hline 
$^3$P$_0$ &  5 & 6 & 7.9 & 9.7 & 0.8 & 12.3\\ \hline 
$^3$P$_1$ &  17 & 4 & 16.9 & 27.0 & 1.9 & 26.1\\ \hline 
$^3$P$_2$ &  16.5 & 0.7 & 16.0 & 27.7 & 1.2 & 25.8\\ \hline 
$\epsilon\left(1^-\right)$ &  -10 & 20 & -8.9 & -12.2 & 1.7 & -8.3 \\ \hline 
\end{tabular}
\label{globalphasetable}
\end{center}
\end{table}

\begin{table}
\begin{center}
\caption{Global phase-shift analysis results at 4.00 and 5.55 MeV} 
\begin{tabular}{ | c | r@{$\pm$}l | c | r@{$\pm$}l | c |}
\hline
 & \multicolumn{3}{c |}{4.00 MeV}& \multicolumn{3}{c |}{5.55 MeV} \\ \hline
{Phase} & \multicolumn{2}{c|}{Present} & {CD-Bonn}  & \multicolumn{2}{c|}{Present} & {CD-Bonn}\\ \hline
$^1$S$_0$ &  -56.3 & 0.6 & -56.8 & -67.8 & 0.9 & -67.1\\ \hline 
$^3$S$_1$ &  -49.3 & 0.5 & -49.7 & -58.6 & 0.3 & -59.2\\ \hline 
$^1$P$_1$ &  17.3 & 1.6 & 18.2 & 21.2 & 1.7 & 22.5\\ \hline 
$^3$P$_0$ &  14.1 &0.9 & 16.6 & 21.3 & 0.7 & 23.9\\ \hline 
$^3$P$_1$ &  34.9 & 0.3 & 33.9 & 45.2 & 0.5 & 43.0\\ \hline 
$^3$P$_2$ &  37.6 & 0.6 & 34.9 & 51.5 & 0.5 & 47.0\\ \hline 
$\epsilon\left(1^-\right)$ &  -13 & 2 & -9.0 & -14 & 2 & -9.6 \\ \hline 
\end{tabular}
\label{globalphasetable2}
\end{center}
\end{table}

\section{Comparison with Theoretical Calculations}

We first compare our new experimental results with recent \textit{ab initio} momentum-space calculations from Deltuva and Fonseca \cite{Del07b} which rigorously include the Coloumb interaction and use a variety of 2N potentials.  For simplicity, only their results obtained using the CD-Bonn potential \cite{Mac01} are shown in Figures \ref{a0y}-\ref{axx}, but the results of the other realistic 2N potentials considered, AV18 \cite{Wir95} and N3LO \cite{Ent03}, are similar.  The A$_{0y}$ calculations consistently underpredict the new results by 10-20\% at the maximum.  This is similar to, though smaller than, the previously established 40\% underprediction of A$_{y0}$ by several realistic potentials and theoretical methods \cite{Del07b, Fis06}.  

The theoretical results for the spin-correlation coefficients at E$_{p}$ = 2.77 MeV and above agree with the present results at backward angles but are too small by about 0.02 at forward angles.  The disagreement for backward angles between the theoretical results and the present phase-shift analysis at 2.28 MeV may result from the lack of back-angle data points, especially for A$_{xx}$.

\begin{figure}[h]
\begin{center}
\includegraphics[width= 4in]{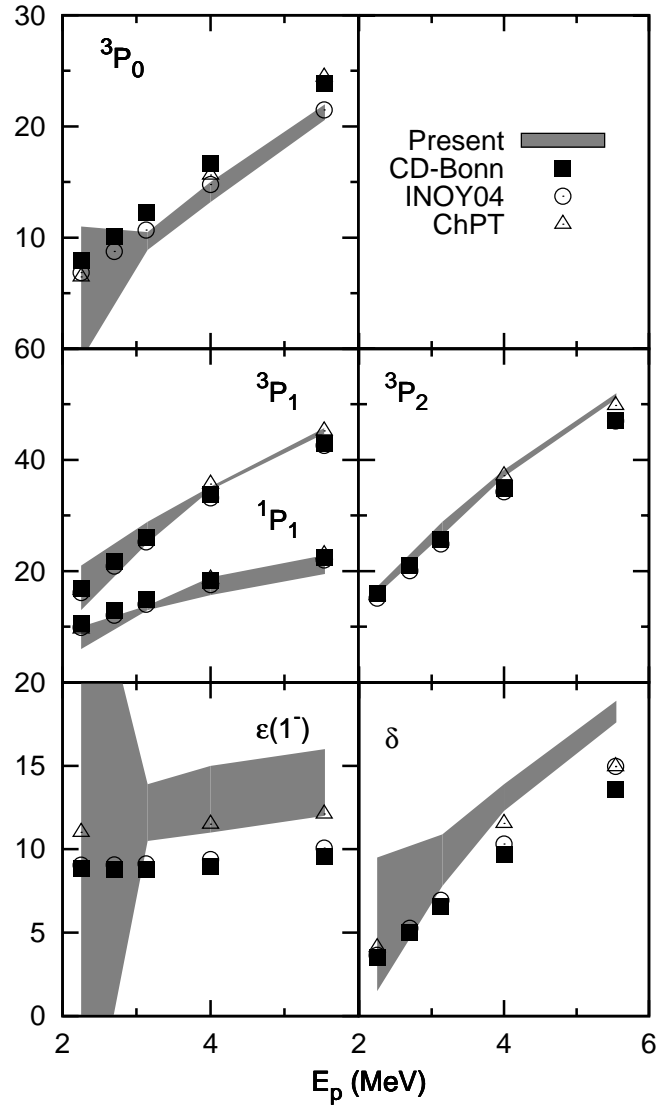}
\caption{(Color online) Comparison of the present p-wave phase-shifts,  in degrees, with the theoretical results of Deltuva and Fonseca \cite{Del07b} using the CD-Bonn and Doleschall INOY04 potentials and of Viviani \cite{Viv10} using a potential derived from chiral potential theory (ChPT).  The splitting $\delta$ is defined in the text.}
\label{pwaves}
\end{center}
\end{figure}

For the phase-shifts, the theoretical S-waves generally agree well with the present results, while the theoretical $^1$P$_1$ and $^3$P$_0$ phase-shifts are larger.  The theoretical  $^3$P$_1$ and $^3$P$_2$ phase-shifts, as well as the $\epsilon$(1$^{-}$) mixing parameter, are consistently smaller than our present results.  The splitting between the triplet P-waves is also underpredicted, as shown in Fig. \ref{pwaves}, where $\delta$ = $^3$P$_2$ - ( $^3$P$_1$ +  $^3$P$_0$)/2.  This is interesting in light of the strong dependence of A$_{y0}$ on that splitting \cite{Fis06}.

\begin{figure}[h]
\begin{center}
\includegraphics[width=2.8in]{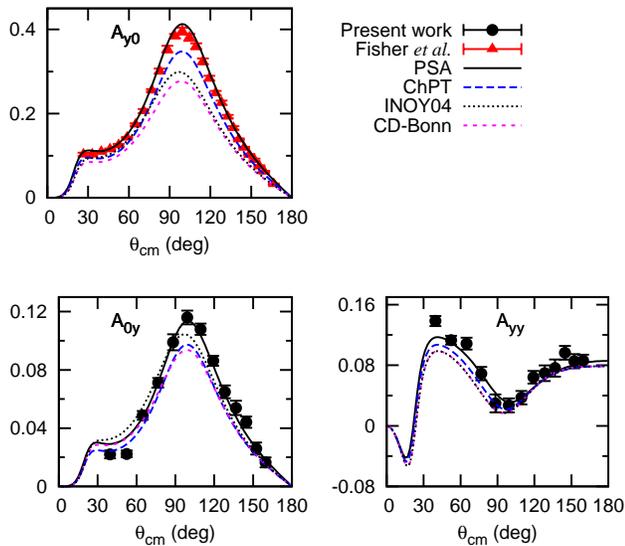}
\caption{(Color online) Comparison of the present observables A$_{0y}$ and A$_{yy}$ measured at 4 MeV, along with the A$_{y0}$ measurements of Fisher \textit{et al.} \cite{Fis06}, with the theoretical results of Viviani \cite{Viv10} using a using a potential derived from chiral potential theory (ChPT).  The results of Deltuva and Fonseca  \cite{Del07b} using the CD-Bonn and INOY04 potentials are also shown. }\label{theory_obs}
\end{center}
\end{figure}

Results from Deltuva and Fonseca using the Doleschall potential INOY04 \cite{Dol04} are also shown in Figures \ref{pwaves} and \ref{theory_obs}.  That potential introduces non-localities to simulate implicitly the effect of three-nucleon forces, which are necessary to reproduce three-and four-nucleon binding energies.  The parameters of those non-localities are adjusted to reproduce 3N scattering phase-shifts, and also better to reproduce the A$_{y0}$ measurements of Ref. \cite{Fis06}.  Here, this potential improves the description of A$_{0y}$ somewhat, but has little effect on A$_{xx}$ and A$_{yy}$.  Considering the phase-shifts, this model better describes $^3$P$_0$ and produces a P-wave splitting closer to the experimental results. 

The addition of explicit phenomenological 3N forces in theoretical calculations has traditionally not provided full agreement between experiment and theory for p + $^3$He observables, especially for A$_{y0}$.  New results from Viviani \cite{Viv09, Viv10} using a 2N \cite{Ent03} and 3N \cite{Ber08} interaction derived from chiral perturbation theory at N3LO and N2LO, respectively, are shown in Fig. \ref{theory_obs}.  The calculations, made using the Kohn variational principle and the hyperspherical harmonic technique, are compared with both the present A$_{0y}$ and A$_{yy}$ results and A$_{y0}$ from Ref. \cite{Fis06} at 4 MeV.  Satisfying reduction of ÔA$_y$ puzzleÕ differences was obtained using this effective-field-theory version of the 3N interaction, corresponding to the improved agreement with the triplet P-wave phase-shifts shown in Fig. \ref{pwaves}.  Better agreement for the $\epsilon(1^-)$ mixing parameter is also evident.  Though theoretical agreement with experimental results is still not complete, Machleidt has suggested \cite{Mac09} that sizeable one-loop 3N force diagrams exist at N4LO of the $\Delta$-less chiral theory, or at N3LO when a phenomenological $\Delta$ is included, and that their addition may ultimately explain the remaining differences. 

\section{Conclusions}
In this paper we have presented new measurements of A$_{0y}$, A$_{xx}$, and A$_{yy}$ for p+$^3$He elastic scattering between 2 and 6 MeV proton energy.  The target analyzing power measurements represent an improvement in accuracy over previous results, while the spin-correlation measurements include the lowest-energy data to date. 

These new measurements were included in a global phase-shift analysis and new phase-shifts were extracted.  These additional data remove the ambiguity reported in Ref. \cite{Geo03}.  Though discontinuities in the energy-dependence of the D- and F-wave phase-shifts were present, a single global solution was obtained by requiring that all phases be continuous over the energy range of the global database.  

Recent theoretical calculations \cite{Del07b} using realistic 2N potentials underpredict the present A$_{0y}$ results by 10-20\%, which is similar to but smaller than the previously observed 40\% underprediction of A$_{y0}$ \cite{Del07b,Fis06}.  The spin-correlation coefficients A$_{xx}$ and A$_{yy}$ are better-described, though small underpredictions are observed at forward angles, and qualitatively different trends are observed at 2.25 MeV.  The S-wave phase shifts agree well, but $^1$P$_1$, $^3$P$_0$, and $\epsilon\left(1^-\right)$ differ, and the theoretical triplet P-wave splitting is too small.  

The INOY04 potential \cite{Dol04} which includes a phenomenological 3N force improves the description of A$_{0y}$ and $^3$P$_0$ and increases the P-wave splitting.  Preliminary calculations by Viviani \cite{Viv09} of p + $^3$He elastic scattering observables at low energies using chiral 2N and 3N potentials show closer agreement with A$_{y0}$.

\section{Acknowledgements}
The authors gratefully acknowledge continuing theoretical support for this work from M. Viviani (Pisa) and A. Deltuva and A. Fonseca (Lisbon). We wish to thank E. George for making available to us the energy-dependent p+$^3$He phase-shift analysis code used previously, and J. Dunham and B. Carlin whose continuing technical support throughout the project was essential to its successful completion.  Additionally, we thank M. Boswell, C. Angell, and J. Newton for assistance with data-taking and R. Prior for supplying the p+$^4$He phase-shift code.  Financial support from the US Department of Energy Office of Nuclear Physics under Grant \# DE-FG02-97ER4041 is also gratefully acknowledged.

\end{document}